\documentclass[9pt,sigconf]{acmart}
\AtBeginDocument{%
  \providecommand\BibTeX{{%
    \normalfont B\kern-0.5em{\scshape i\kern-0.25em b}\kern-0.8em\TeX}}}

\usepackage[english]{babel}
\usepackage{blindtext}
\usepackage{siunitx}
\usepackage{graphicx}
\usepackage{subcaption}
\usepackage{balance}

\usepackage{algorithm}
\usepackage[noend]{algpseudocode}

\usepackage{url}
\usepackage{tablefootnote}
\usepackage{booktabs}
\usepackage{multirow}
\usepackage{varwidth}

\setcopyright{none}
\renewcommand\footnotetextcopyrightpermission[1]{}

\acmBooktitle{Workshop on Network and Operating System Support for Digital Audio and Video (NOSSDAV '21)) (NOSSDAV '21), September 28-October 1, 2021, Istanbul, Turkey}
\acmDOI{10.1145/3458306.3460998}

\copyrightyear{2021}
\acmYear{2021}
\setcopyright{acmcopyright}\acmConference[NOSSDAV '21]{Workshop on Network and Operating System Support for Digital Audio and Video (NOSSDAV '21))}{Sep. 28-Oct. 1, 2021}{Istanbul, Turkey}
\acmBooktitle{Workshop on Network and Operating System Support for Digital Audio and Video (NOSSDAV '21)) (NOSSDAV '21), September 28-October 1, 2021, Istanbul, Turkey}
\acmPrice{15.00}
\acmDOI{10.1145/3458306.3460998}
\acmISBN{978-1-4503-8435-3/21/09}

\newcommand{\model}{\textsf{360NorVic}}

\begin{document}

\title{\model: 360-Degree Video Classification from Mobile Encrypted Video Traffic}

\author{Chamara Kattadige}
\email{ckat9988@uni.sydney.edu.au}
\affiliation{The University of Sydney}

\author{Aravindh Raman}
\email{aravindh.raman@telefonica.com}
\affiliation{Telefonica Research}

\author{Kanchana Thilakarathna}
\email{kanchana.thilakarathna@sydney.edu.au}
\affiliation{The University of Sydney}

\author{Andra Lutu}
\email{andra.lutu@telefonica.com}
\affiliation{Telefonica Research}

\author{Diego Perino}
\email{diego.perino@telefonica.com}
\affiliation{Telefonica Research}

\renewcommand{\shortauthors}{chamara, et al.}

\begin{abstract}
   Streaming \ang{360} video demands high bandwidth and low latency, and poses significant challenges to Internet Service Providers (ISPs) and Mobile Network Operators (MNOs). The identification of \ang{360} video traffic can therefore benefits fixed and mobile carriers to optimize their network and provide better Quality of Experience (QoE) to the user. However, end-to-end encryption of network traffic has obstructed identifying those \ang{360} videos from regular videos. As a solution this paper presents \model, a near-realtime and offline Machine Learning (ML) classification engine to distinguish \ang{360} videos from regular videos when streamed from mobile devices. We collect packet and flow level data for over 800 video traces from YouTube \& Facebook  accounting for 200 unique videos under varying streaming conditions. Our results show that for near-realtime and offline classification at packet level, average accuracy exceeds 95\%,  and that for flow level, \model{} achieves more than 92\% average accuracy. Finally, we pilot our solution in the commercial network of a large MNO showing the feasibility and effectiveness of \model~in production settings.
\end{abstract}



\maketitle

\section{Introduction}

Mobile video streaming has been dominating the data traffic growth over the years accounting for over 75\% of global data traffic in 2020~\cite{sillhouette}. 
Recent advancements in networking and multimedia technologies have paved the way for pervasive accessibility of immersive video streaming, a.k.a. \ang{360} videos.
Online streaming of immersive content is increasingly getting popular, and is now widely available on popular video streaming platforms such as YouTube (YT) and Facebook (FB)~\cite{online_YT,online_FB}.

Unlike conventional videos, \ang{360} videos offer a truly omnidirectional view that gives users the freedom to interact with the video content.
However, streaming \ang{360} videos over mobile networks poses significant challenges. First, \ang{360} videos demand up to 80x more bandwidth to achieve the same user-perceived quality level of a conventional video \cite{guan2019pano}. Second, only a slight delay (just above 25ms) between the perception of an action and image display will make the user suffer cyber- or motion-sickness~\cite{liu2018cutting}. 
Hence, wide adaptation of \ang{360} video asserts a significant strain on the network.

To this end, identification of \ang{360} video traffic generated from mobile devices significantly benefits ISPs and MNOs in network capacity planning and traffic optimizations, such as to maintain traffic shaping (prioritizing and balancing), traffic policing effectively \cite{Meauring_video_illias}. Furthermore, aggregated results such as, frequency and geographical usage of \ang{360} videos, can provide information to facilitate new caching, compression, transcoding strategies as well as dynamic radio resource allocation in the case of cellular networks. Hence, having required and sufficient resources and new streaming strategies will guarantee a smooth streaming (i.e., low re-buffering and quality changes) of high quality \ang{360} videos while increasing the end user Quality of Experience (QoE).

Traditional, deep packet inspection (DPI) is becoming ineffective due to wider adoption of end-to-end encryption. In fact, the majority of video traffic of today, if not all, are encrypted as they leverage HTTPS/TCP or QUIC/UDP protocols. In light of this, there have been significant efforts in encrypted traffic classification \cite{Meauring_video_illias,sillhouette,deepcontent,end2end_encrypted_traffic_classification,lotfollahi2020deep}. In particular, recent work shows that accurate identification of video flows is possible, despite the content being transferred is encrypted \cite{deepcontent}. Nevertheless, none of these classification methods supports extraction or identification of \ang{360} video traffic out of normal video traffic. 

In this paper, we propose \model: \textbf{360} \& \textbf{Nor}mal \textbf{Vi}deo \textbf{C}lassification engine that takes encrypted packet or flow level data generated by mobile devices as inputs.
\emph{To the best of our knowledge, this is the first study that demonstrates the feasibility of classifying \ang{360} videos from conventional encrypted video traffic.}

We develop a robust classification engine based on XGBoost relying on the accessible features from the encrypted data such as throughput, packet size or packet number. We make accurate identification of \ang{360} video traffic both in near real-time and offline. We train and evaluate \model~with video traffic data that we capture while streaming videos on smartphones from YT and FB, the two most popular \ang{360} video service providers, who leverage  GQUIC/UDP and HTTPS/TCP protocols respectively. 
Depending on the vantage point of interception, the 
information available in network traffic data vary significantly. 
For example, intercepted traffic at the local router or at the user device consists with packet level data whereas, at the ISP or MNO level,  flow-level data is prominent. We collected over 800  video traces from over 200 unique video streams under varying conditions. We achieve 95\% and 92\% offline classification accuracy at packet level and  flow level respectively.
Moreover, near real-time classification exceeds 95\% accuracy after 25 seconds of data.
Artifacts of this work (i.e., dataset and tools) are available at \url{https://github.com/manojMadarasingha/360norvic.git}

Finally, to demonstrate the capability of \model~and how it can interact with real-world middle-boxes operating in a mobile network environment, we also run a controlled campaign to pilot our technology in an operational MNO. 
For this, we capture flow-level statistics for a set of controlled video traffic traces at the MNO level. The pilot results are in line with the ones obtained in the experimental testbed showing the feasibility and effectiveness of \model~in production settings. 

\begin{figure*}[t!]
    \centering
    \captionsetup{justification=centering}
    
    \begin{subfigure}{.24\textwidth}
        \centering
        \includegraphics[width=\linewidth]{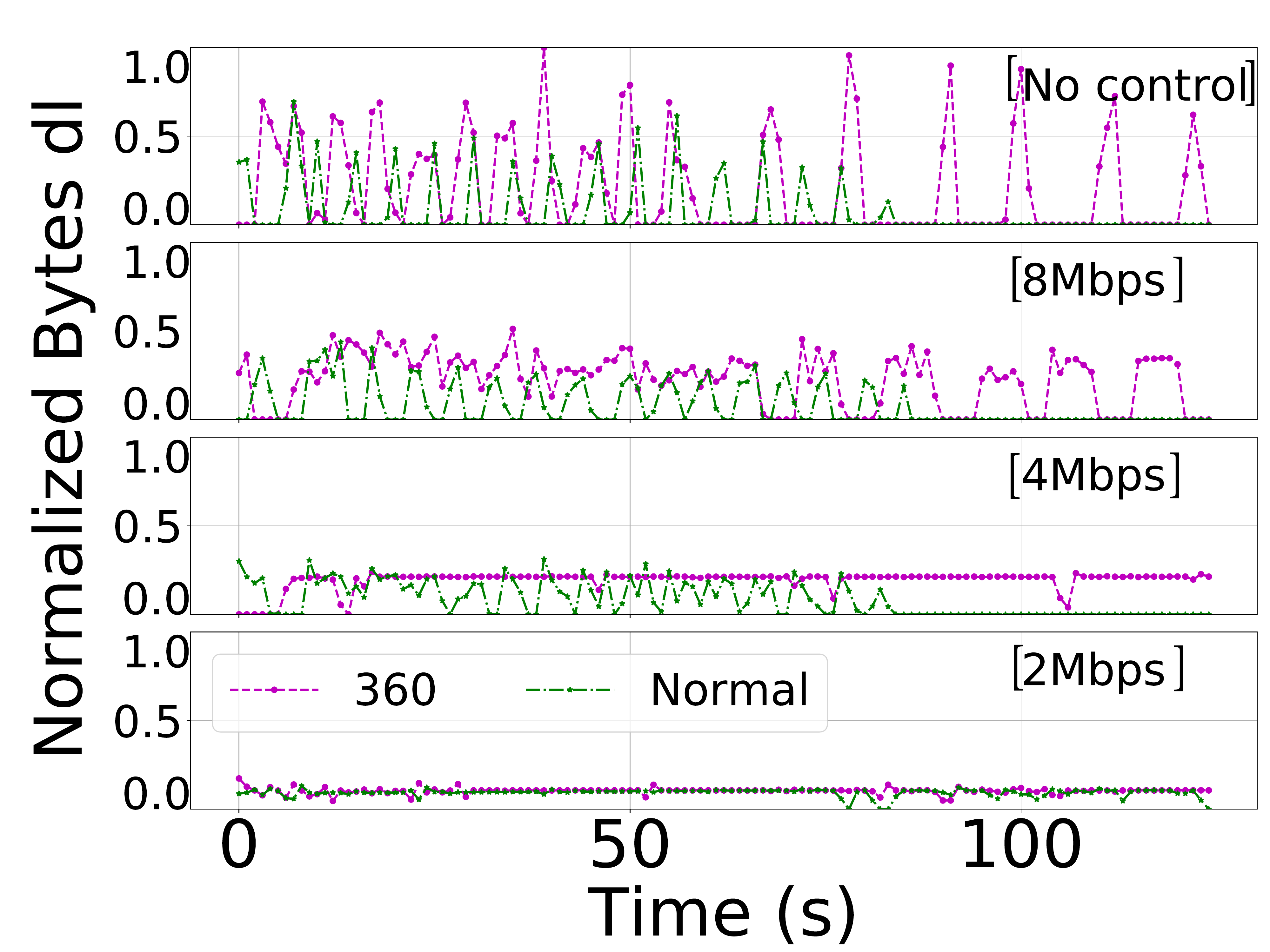}\vspace{-2mm}
        \caption{Bandwidth impact (YT)}
        \label{figure:impact from bw}
    \end{subfigure}    
    \hfill
    \begin{subfigure}{.37\textwidth}
        \centering
        \includegraphics[width=\linewidth]{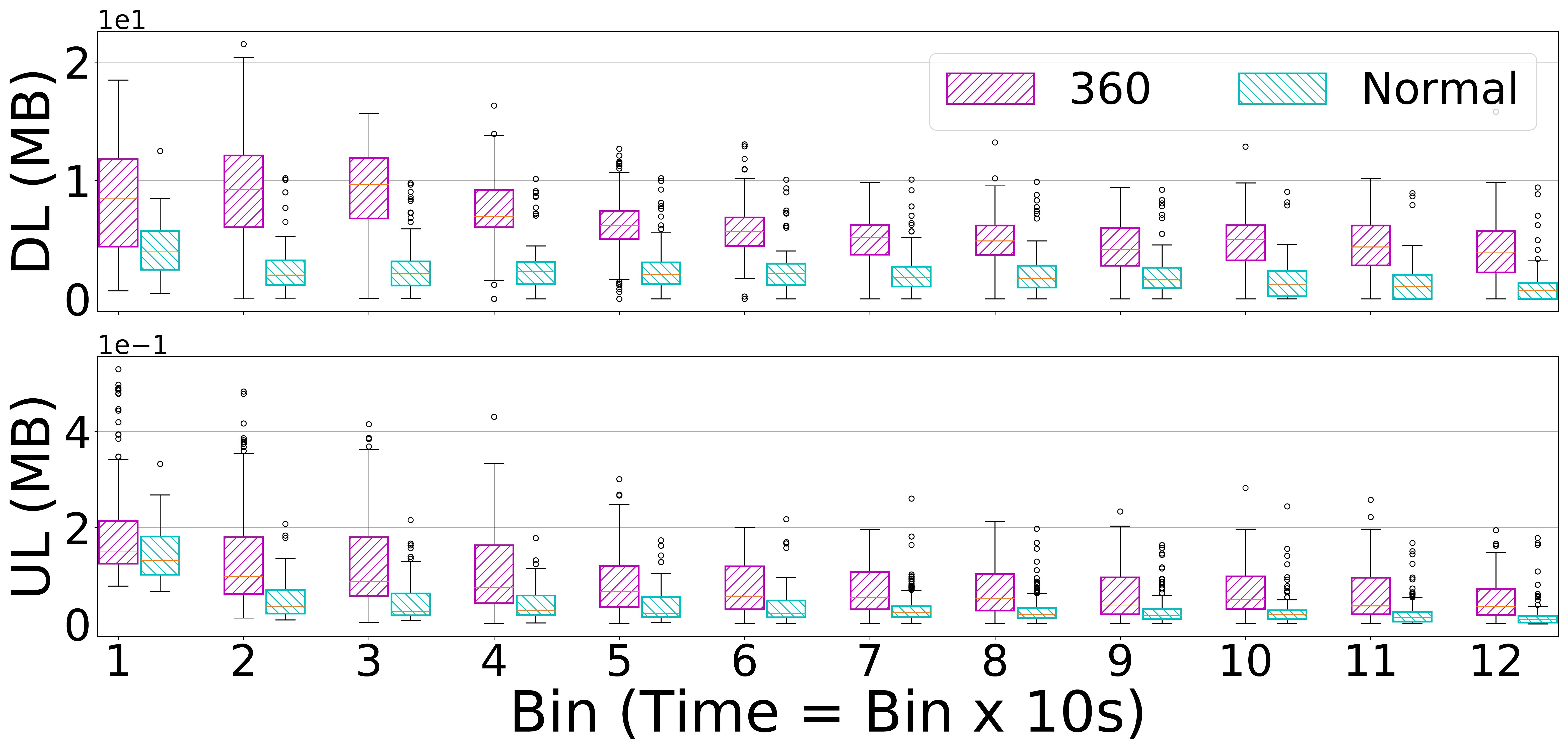}\vspace{-2mm}
        \caption{Data: DL ($\small \times 10^{1}$MB) \& UL ($\small \times 10^{-1}$MB) for YT}
        \label{figure:data dl_ul yt}
    \end{subfigure}
    \hfill
    \begin{subfigure}{.37\textwidth}
        \centering
        \includegraphics[width=\linewidth]{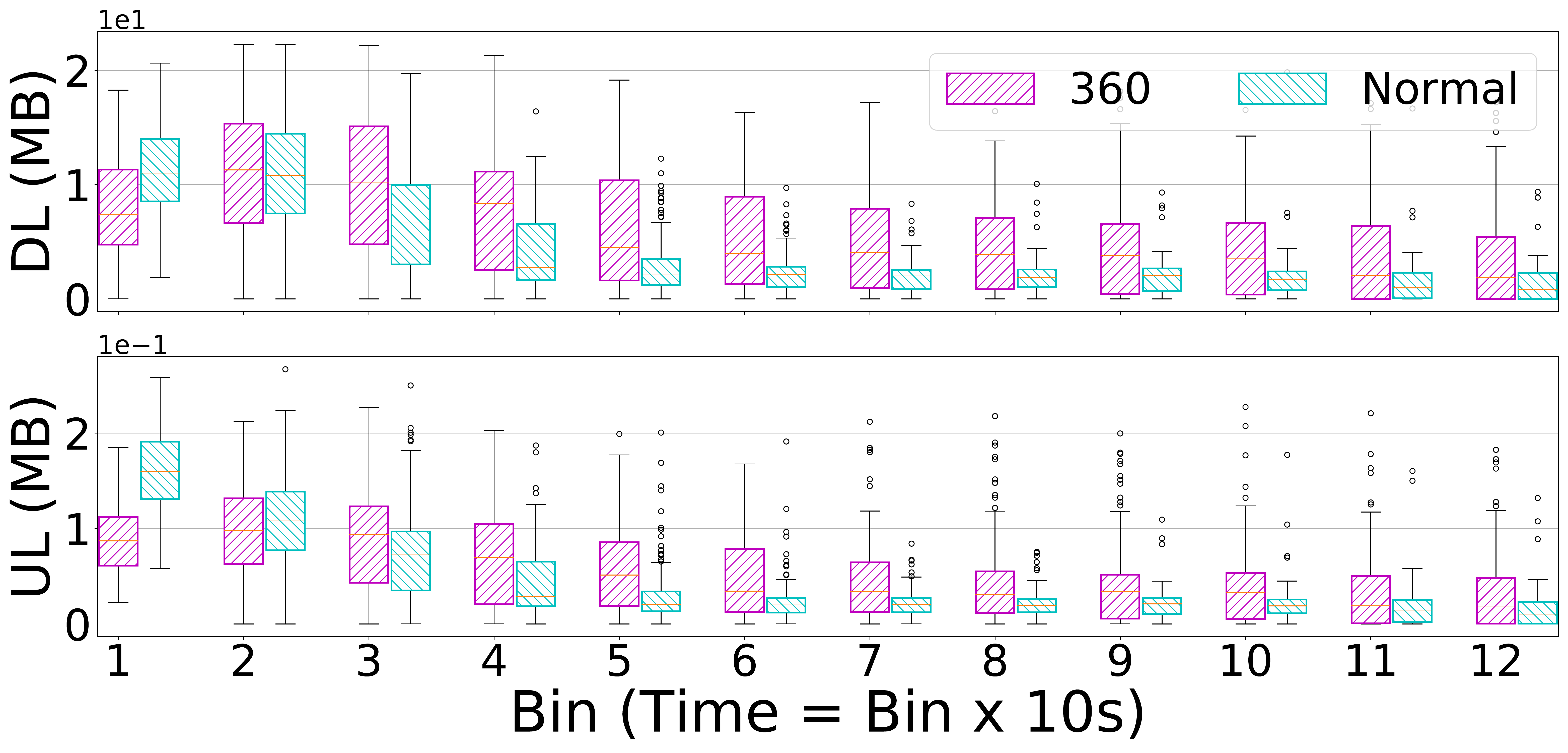}\vspace{-2mm}
        \caption{Data: DL ($\small \times 10^{1}$MB) \& UL ($\small \times 10^{-1}$MB) for FB}
        \label{figure:data dl_ul fb}
    \end{subfigure}\vspace{-2mm}
    \caption{Bandwidth impact, DL (Downlink) and UL (Uplink) patterns for 10s of binned data: YT and FB  (\ref{figure:data dl_ul yt}~\&~\ref{figure:data dl_ul fb})
    }
    \label{figure: packet level realtime performance}
\end{figure*}

\section{Related Work}
\label{sec:related_work}

Classification of video flows out of encrypted traffic has been well-studied 
including video flow classification \cite{sillhouette, who_is_the_king}, identifying exact video \cite{deepcontent, walls_have_ears}, and QoE of video being transferred \cite{inferring_streaming_video_quality,Meauring_video_illias, a_framework_for_qoe_analysis,requet}.  
In \cite{sillhouette}, a 
flow characteristics were leveraged to identify YT flows, instead of Machine Learning (ML) based classifiers. 
Gu \textit{et al.} \cite{walls_have_ears} suggest that using encrypted traffic, videos can be re-identified uniquely, observing bit-rate trends.
A study by Li \textit{et al.} \cite{deepcontent} fingerprints videos using frame level data passively extracted by sniffing a WiFi network training a Recurrent Neural Network (RNN) and Multi Layer Perception (MLP)  network gaining 97\% of accuracy. 

In recent studies, efficient mechanisms to deliver  \ang{360} videos such as viewport (VP) adaptive streaming~\cite{flare,rubiks,poi360}, VP prediction and encoding \cite{optile} have been proposed.
Diversely, LIME~\cite{lime} live streams \ang{360} videos using both crowd-source and lab (controlled) setups on platforms YT and FB, showing that existing networks cannot provide the desired \ang{360} video QoE in live streaming.
A similar study done by Yi \textit{et al.} \cite{measurement_study_nossdav} focused on \ang{360} 4K live streaming, and shows that streaming \ang{360} video in 480p to 4K resolution can impose 13s to 42s delay. 
Afzal \textit{et al.}~\cite{360_video_characterization} explore 2285 YouTube \ang{360} videos comparing the video characteristics such as duration, resolution and bitrate with normal videos showing that even if the bitrate of \ang{360} video is greater than normal video, bitrate variability is comparably low, which can benefit the network provisioning.
\emph{Despite the plethora of work in video flow classification and efficient \ang{360} video delivery, a little has been done on classification of \ang{360} video traffic in real networks, further motivating our work to extract \ang{360} video flows out of encrypted video  traffic
}

\section{Background and Challenges}
\label{sec:motivation}

\ang{360} videos differ in many aspects (e.g. encoding \& decoding, bandwidth requirement) from normal videos. However, our experimental analysis based on HTTP Archive Format (HAR) data (i.e., collected by streaming videos for YT/FB \& \ang{360}/Normal) shows that there is no difference in protocols used for \ang{360} and normal video streaming and in user-interactiveness with \ang{360} videos. This analysis shows that both platforms still stream the entire panoramic video frame despite the benefits of VP adaptive streaming~\cite{flare,rubiks,poi360}. 

Figure~\ref{figure:impact from bw} shows the bytes downloaded (dl) at packet level for videos streamed under different bandwidth conditions: No-control (NC: without controlling bandwidth), 8, 4 and 2 Mbps.
Values are max normalized, taking the maximum bytes dl  value in NC condition. In all conditions, \ang{360} videos streamed more data than normal videos. 
In all other conditions except 2Mbps, \ang{360} videos buffer data until the end of the video, whereas normal videos finished downloading before the end of the video. Figure~\ref{figure:data dl_ul yt}~and~\ref{figure:data dl_ul fb} show the data dl and uploaded (ul) distributions for all the traces aggregated to 10s bin.
In addition to the higher data transmission by \ang{360} videos, there are differences in streaming between two content platforms, due to the differences in ABR algorithms, mobile apps etc. (i.e., compared to \ang{360} videos, YT streams normal videos at a steady pace, but FB streams more data for normal videos at the beginning.)

Despite the noticeable differences in traffic patterns between \ang{360} and normal videos, it is still challenging to classify them due to two main reasons. First, under different bandwidth conditions both types of videos can show similar streaming patterns (i.e., at beginning of the video, lower bandwidth values), which makes it difficult for real-time classification. Second, from Figure~\ref{figure:data dl_ul yt}~and~\ref{figure:data dl_ul fb}, despite the clear difference between 25\textsuperscript{th} and 75\textsuperscript{th} percentile ranges of \ang{360} and normal videos, considering the entire range, there are significant overlaps. One reason is that, at the beginning, players buffer more data irrespective of the video type. Similarly, towards the end, data dl~\&~ul gradually decrease making buffered data volume of both video types closer to each other~\cite{schwind2020dissecting}. Therefore, leveraging features such as entire bytes/packet downlink/uplink would give sub-optimal results.

\section{Methodology}\label{sec:methodology}

Now we present \model, an online and offline \ang{360} video classification platform taking the encrypted network traffic generated by mobile devices as the input. Figure \ref{figure:360_norvic_overview} shows overall work flow of \model, including the locations where its input data can be collected (\textcircled{\footnotesize{1}}, \textcircled{\footnotesize{2}},\textcircled{\footnotesize{3}}). 

\subsection{Dataset}\label{subsection: dataset}
 
\begin{figure}[t]
    \centering
    \includegraphics[width=\linewidth]{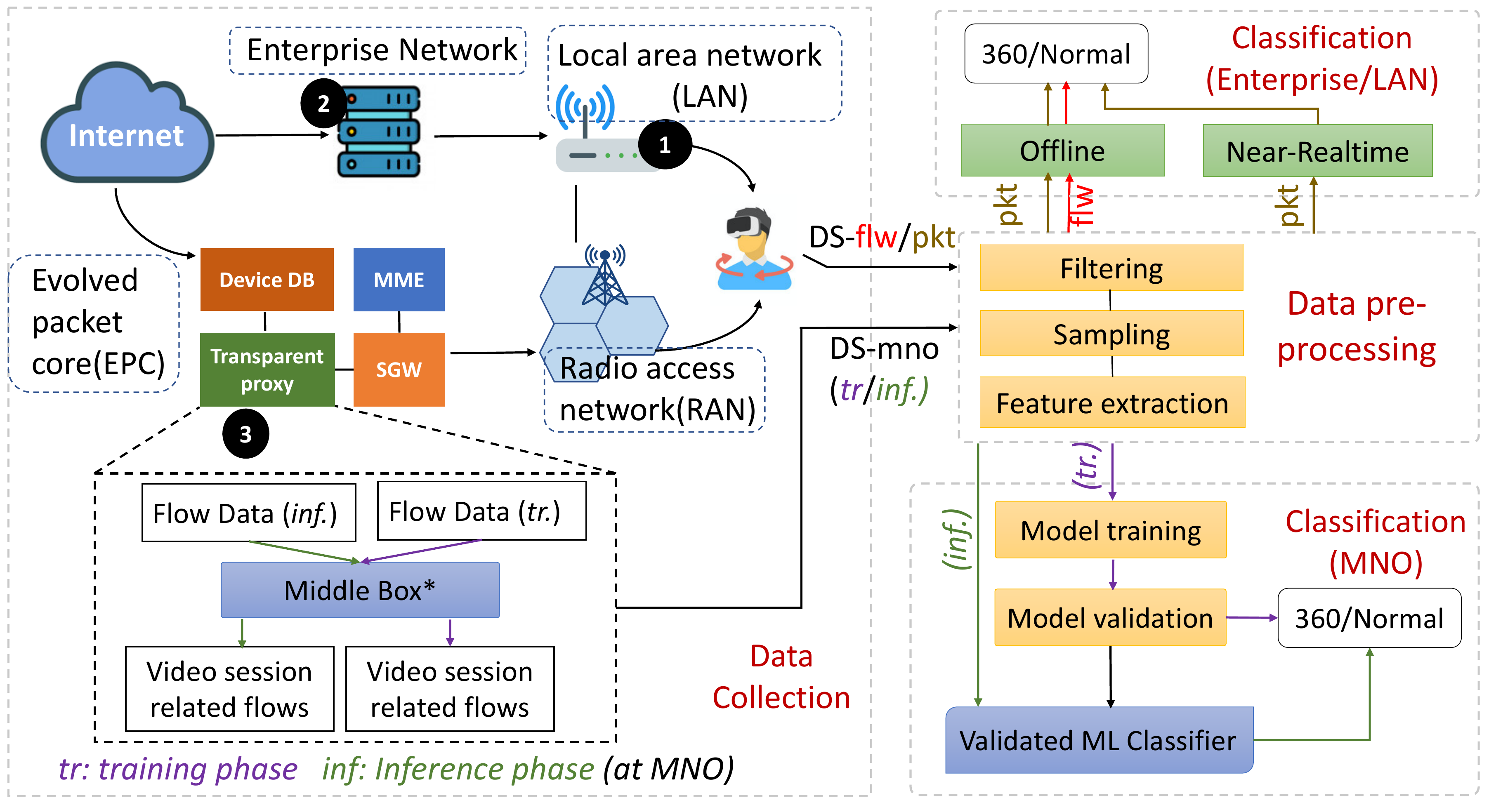}\vspace{-4mm}
  \caption{360NorViC model: Data collection, pre-processing and classification for LAN\textcircled{\footnotesize{1}}, Enterp.\textcircled{\footnotesize{2}} \& MNO\textcircled{\footnotesize{3}} data.}
  \label{figure:360_norvic_overview}
\end{figure}

\textbf{Video selection.} We selected 200 unique videos under four main categories, YT/FB and \ang{360}/Normal
(50 for each category, playback time 2-4 min).
We selected the videos according to their popularity in video distribution channels in YT \cite{ytVR,ytTrend} and FB pages \cite{Mostpopular360vidFB,360videosFB,vtrnd,unilad}, whilst matching the genre\footnote{Selected genres include but not limited to sports, documentary Horror, animals, cartoon, driving, movie trailer, roller coaster, scenery etc.}
of videos among four categories. We focused on \emph{on-demand} videos  due to its popularity compared to live streaming, especially in the case of \ang{360} video live streaming due to the requirement of specialised cameras to record \ang{360} videos. 
\noindent
\textbf{Data collection.}
We collect over 600 encrypted packet traces\footnote{over 200 traces (see Section~\ref{sec:pilot}) at MNO causing over 800 traces in total.} (\textsf{DS-pkt}) at the local router (\textcircled
{\footnotesize{1}}/PC in our experiment setup, cf. Fig.~\ref{figure:360_norvic_overview}), while streaming the selected videos with the latest versions of YT/FB apps.
We also vary the streaming context conditions, as follows:
(i) Time of the day: morning, afternoon and night/evening, 
(ii) Bandwidth: Not Controlled (\textbf{NC}-sufficient bandwidth for streaming), and 
(iii) Backbone network: 802.11n with wired broadband, and 4G/LTE. 
Due to the Adaptive Bit-Rate (ABR) streaming and
\textbf{NC} bandwidth scenario, resolution of the videos can vary between 480p--1080p.
                                       
        
        
Using \textsf{DS-pkt}, we synthesize \textsf{DS-flw}, which includes traffic flows emulating data collection at vantage point \textcircled{\footnotesize{2}}. 
\textsf{DS-flw} is compatible with the data formats deployed by ISPs in their systems.
YT videos are primarily streamed with QUIC protocol while FB leverages UDP protocol.
However, due to many other non-video flows (e.g., audio streams, advertisements etc.), all three types of packets (i.e., including TCP) are observed in collected traces.

\vspace{-1mm}
\subsection{Data Pre-processing}\label{subsection: data pre processing}

\textbf{Packet level (\textsf{DS-pkt}).}
We first filter packets related to the user device using the MAC address. 
We consider bins resulting from a 5s sliding window (by 1s steps) for calculating the features we list in Table \ref{table: features_packet_level} taking insights from literature ~\cite{deepcontent}\cite{end2end_encrypted_traffic_classification}\cite{inferring_streaming_video_quality}.
\begin{table}[h!]
\centering. 
\captionsetup{justification=centering}
\small{
\caption{Features extracted at packet level}\vspace{-4mm}
\label{table: features_packet_level}
\begin{tabular}{p{2cm}p{4.5cm}} 
    \toprule
    \textbf{Direction} & \textbf{Feature} \\
    \midrule
    Uplink (ul)     & Frame length\tablefootnote{Frame length on the wire} in bytes (Total)    \\ 
    Downlink (dl)   & Packet size\tablefootnote{Total length of the packet from transport layer to application layer } in bytes (Total)       \\
                    & TCP-header length in bytes (Total)      \\
                    & Num of packets (Total)      \\
                    & Packet size (mean, min,  max, std) \\ 
    \bottomrule
\end{tabular}
}
\end{table}
We take these bin level features for our near real-time classification. For the offline prediction model, we calculate the statistical features, mean, std, min, max, (25$^{th}$, 50$^{th}$ and 75$^{th}$) percentile for each of these features over the bins covering the first \textit{t} seconds 

\noindent
\textbf{Flow level (\textsf{DS-flw}).} We filter flows for individual users based on the MAC address,
A flow is defined using 5 parameters: source/destination 
IP/port and the protocol. To mitigate the impact of non-video flows such as background traffic, analytics, advertisements etc., we sample the flows related to YT and FB observing the domain specific keywords, i.e. YT: \textit{googlevideo, yt, youtube}, FB: \textit{fb, fbcdn, facebook} \cite{sillhouette,Meauring_video_illias}.
We extract the features in Table \ref{table:features flow level}, for ul \& dl for the entire video flow. Then, mean, sum, min, max, of each features are calculated taking first \textit{n} number of flows, sorted by bytes dl value in descending order. When {$n=1$}, we consider only the mean value. 
The reason behind varying the number of flows is to see its impact on classification accuracy. Taking all the flows can add more noise. Nevertheless, leaving some flows out of processing can drop important video streaming information.

\vspace{-3mm}
\begin{table}[h]
\centering 
\small{
\caption{Features extracted at flow level}\vspace{-4mm}
\label{table:features flow level}
\begin{tabular}{p{3.6cm}p{3.8cm}} 
    \toprule
    \textbf{Entire video trace (ul/dl)} & \textbf{Within bursts (ul/dl)} \\
     \midrule
    Throughput (mean)                   & Burst size (max), Burst rate (max)             \\ 
    
    Frame gap (mean)                    &  Burst time (max)  \\
  
    Frame size (mean)                   & Burst num of packets    \\
   
    Packet Re-transmission             &  Burst Gap(mean), duration (mean)  \\
  
   
    \bottomrule 
    \end{tabular}
    }
\end{table}


\vspace{-6mm}
\subsection{Classification}\label{sub section: classification models}

We develop two ML models for offline and near real-time classification. Note that, we assume classification of video data (even if the content is encrypted) is relatively straightforward \cite{end2end_encrypted_traffic_classification,deepcontent,velan2015survey,lotfollahi2020deep}. We validate this via private communication with an operational MNO, confirming they deploy in production middle-boxes for media traffic classification from encrypted traffic~\cite{citrix_media,citrix_media_2}, which is further elaborated
in Section~\ref{sec:pilot}. We train both the models for three traffic types. YT \& FB separately and BOTH (more generalized version combining YT and FB data together) 
taking XGBoost as our main classifier due to the fast performance and high accuracy \cite{xgboost}. Important code implementations are avilable with the artifacts\footnote{\url{https://github.com/manojMadarasingha/360norvic.git}}

\noindent
\textbf{Offline classification model.} Each video trace is represented by one feature vector taking summary statistics of each bin for \textsf{DS-pkt} or flow for \textsf{DS-flw}. 
For each dataset, \textsf{DS-pkt} and for \textsf{DS-flw}, we create three XGBoost classifiers for three traffic types, resulting in 6 offline classifiers in total. 
We select most important features as is seen by the ML models during the initial training with all the features.
To see the minimum data amount for a reliable prediction, we train and test models  for varying data collection intervals (i.e., taking bins from 20s to 120s) for \textsf{DS-pkt}, and for varying number of flows sorted in descending order of bytes dl for \textsf{DS-flw}. There were  20  random train/test  (70\%/30\%)  splits for every analysis.

\begin{figure}[h]
    \centering
    \includegraphics[width=0.9\linewidth]{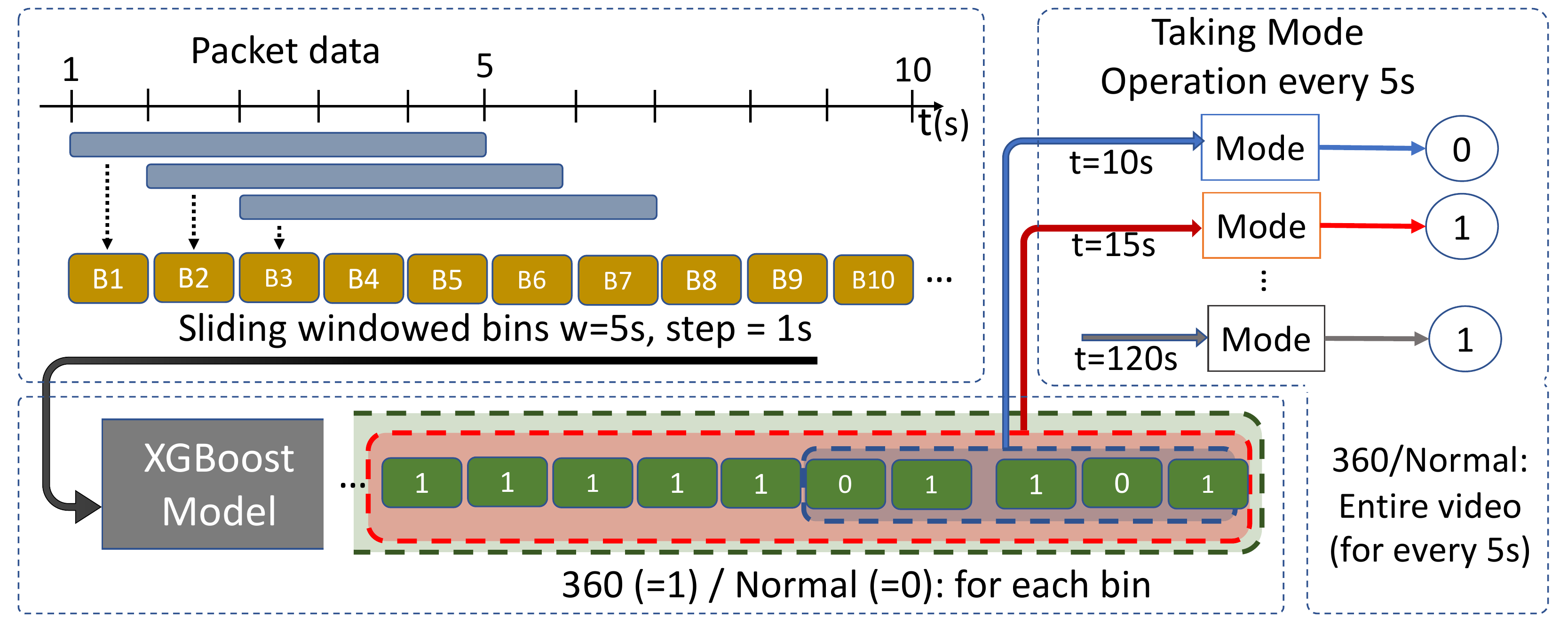}\vspace{-4mm}
  \caption{Near real-time prediction model. XGBoost model takes \textsf{DS-pkt} bins as the input and output 1/0 (\ang{360}/Normal) prediction which is the input for Mode operation}
  \label{figure:real time prediction model}
\end{figure}

\vspace{-3mm}
\noindent
\textbf{Near real-time classification model.} Figure \ref{figure:real time prediction model} shows \model{} near real-time classification model which takes \textsf{DS-pkt} bins individually and output video type (\ang{360}/normal) every 5s with an increasing accuracy until 120s of packet data. 
First, for each bin, XGBoost model predicts whether the bin is from a \ang{360} (=1) or normal (=0) video stream. 
Classification accuracy for these bins separately was comparably lower than \model{} offline results. 
Therefore, to improve the accuracy, we apply \textit{Mode} (majority voting) operation on the output of XGBoost model taking them as groups of binary values (see Figure~\ref{figure:real time prediction model}).  
Every  \textit{n\textsuperscript{th}} \textit{Mode} operation at {$t=(n+1)5s$} takes the first 5\textit{n} bin outputs from the XGBoost model.
We increment group size by 5 bins because it provides traffic data of 10s which can reflect dynamics of video streaming properly \cite{inferring_streaming_video_quality}. Finally, the model (XGBoost$+$Mode) predicts the video class at every 5s.

\section{Results}
\label{sec:Results and analysis}

\subsection{Offline classification}
\label{subsection:results scenario 1 packet_level_off}

\textbf{Packet level (\textsf{DS-pkt}).}
Table \ref{table: performance non realtime packet level} shows the accuracy and F1 score for varying data collection intervals (from 20s to 120s). 
Results show that when we capture packets at least for the first 30s of the video, \model~ achieves accuracy above 93\% (with F1 score $\geq 91\%$) in classifying \ang{360} videos, regardless the app combination (YT, FB or BOTH).
Further increase in interval to 60s slightly increases the accuracy for YT and FB, but drops for BOTH. A further increase to 90s shows that the performance drops for YT,
but remains constant for FB. 
The accuracy further drops at 120s in both cases (YT, FB) because the bins near the end (especially for 2 min videos accounting for 40\% of dataset) of the videos do not contain much data, as those have been already buffered (cf. Figure~\ref{figure:data dl_ul yt}~\&~\ref{figure:data dl_ul fb}). 
Thus, for offline classification, capturing the first 90s data is enough for reasonable accuracy, which saves a large storage capacity at large scale.

\begin{table}[h]
    \centering
    \captionsetup{justification=centering}
    \small{
    \caption{Avg accuracy and F1 score for \textsf{DS-pkt}}\vspace{-4mm}
    \label{table: performance non realtime packet level}
    \begin{tabular}{lSSSSSS}

        \toprule
        {Data collection} &
            \multicolumn{2}{c}{YT(\%)} &
            \multicolumn{2}{c}{FB (\%)} &
            \multicolumn{2}{c}{BOTH (\%)} \\
            interval & {Acc} & {F1} & {Acc} & {F1} & {Acc} & {F1} \\
        \midrule
      
        20s  & 85.9 & 81.9 & 93.7 & 92.7 & 89.8 & 87.2  \\
        30s  & 93.7 & 92.2 & 94.7 & 93.5 & \textbf{95.2} & \textbf{94.1}  \\
        60s  & \textbf{96.7} & \textbf{95.6} & 95.1 & 94.0 & 94.6 & 93.6  \\
        90s  & 95.9 & 94.7 & \textbf{95.1} & \textbf{94.0} & 94.0 & 92.4  \\
        120s & 96.2 & 95.2 & 94.2 & 93.0 & 94.0 & 92.4  \\
        
        \bottomrule
    \end{tabular}
    }
\end{table}

\noindent
\textbf{Flow level (\textsf{DS-flow}).} We present similar results for \textsf{DS-flw} in Table \ref{table: performance offline flow level}, capturing the accuracy and F1 score distribution according to the number of flows we consider for classification. 
For YT, FB and BOTH, taking first 6, 4, and 2 flows gives the best accuracy level respectively. We see similar performance with less flows, which can be highly beneficial for large network operators. 
In fact, taking more flows slightly reduces the performance because it can add more noise to the dataset. 
Differences in inherent streaming properties of YT and FB may have caused low performance in the combined case BOTH. 
In contrast to \textsf{DS-pkt}, flow level accuracy drops by 1-3\% in all three cases. We conjecture this is because flows represent entire streaming sessions, which aggregate the granular features (bin-wise) we considered at packet level. 

\vspace{-3mm}
\begin{table}[h]
    \centering
    \captionsetup{justification=centering}
    \small{
    \caption{Avg accuracy and F1 score for \textsf{DS-flw}}\vspace{-4mm}
    \label{table: performance offline flow level}
    \begin{tabular}{lSSSSSS}
        \toprule
        \multirow{2}{*}{\# of flows}&
            \multicolumn{2}{c}{YT(\%)} &
            \multicolumn{2}{c}{FB (\%)} &
            \multicolumn{2}{c}{BOTH (\%)} \\
            & {Acc} & {F1} & {Acc} & {F1} & {Acc} & {F1} \\
        \midrule
      
        1   & 93.3 & 92.2 & 92.8 & 91.9 & 92.3 & 91.0  \\
        2   & 93.2 & 92.4 & 92.9 & 91.3 & \textbf{92.3} & \textbf{91.1}  \\
        4   & 93.4 & 92.2 & \textbf{94.1} & \textbf{92.8} & 92.2 & 90.8 \\
        6   & \textbf{93.7} & \textbf{92.7} & 92.8 & 91.2 & 91.8 & 90.1  \\
        8   & 92.4 & 90.8 & 91.1 & 88.8 & 89.7 & 87.9  \\
        All & 92.0 & 90.7 & 90.6 & 88.4 & 88.7 & 88.6  \\
        
        \bottomrule
    \end{tabular}
    }
\end{table}

\vspace{-5mm}
\subsection{Near real-time classification}

We perform near-realtime classification for \textsf{DS-pkt}, as it predicts at every 5s interval until accuracy reaches a satisfactory level as shown in Figure \ref{figure: packet level realtime performance}. In addition, it does not require retraining or to maintain multiple models for different lengths of test data.

\begin{figure}[h]
\centering
\vspace{-3mm}
\captionsetup{justification=centering}
   
  \begin{subfigure}{.49\columnwidth}
    \centering
    \includegraphics[width=\linewidth]{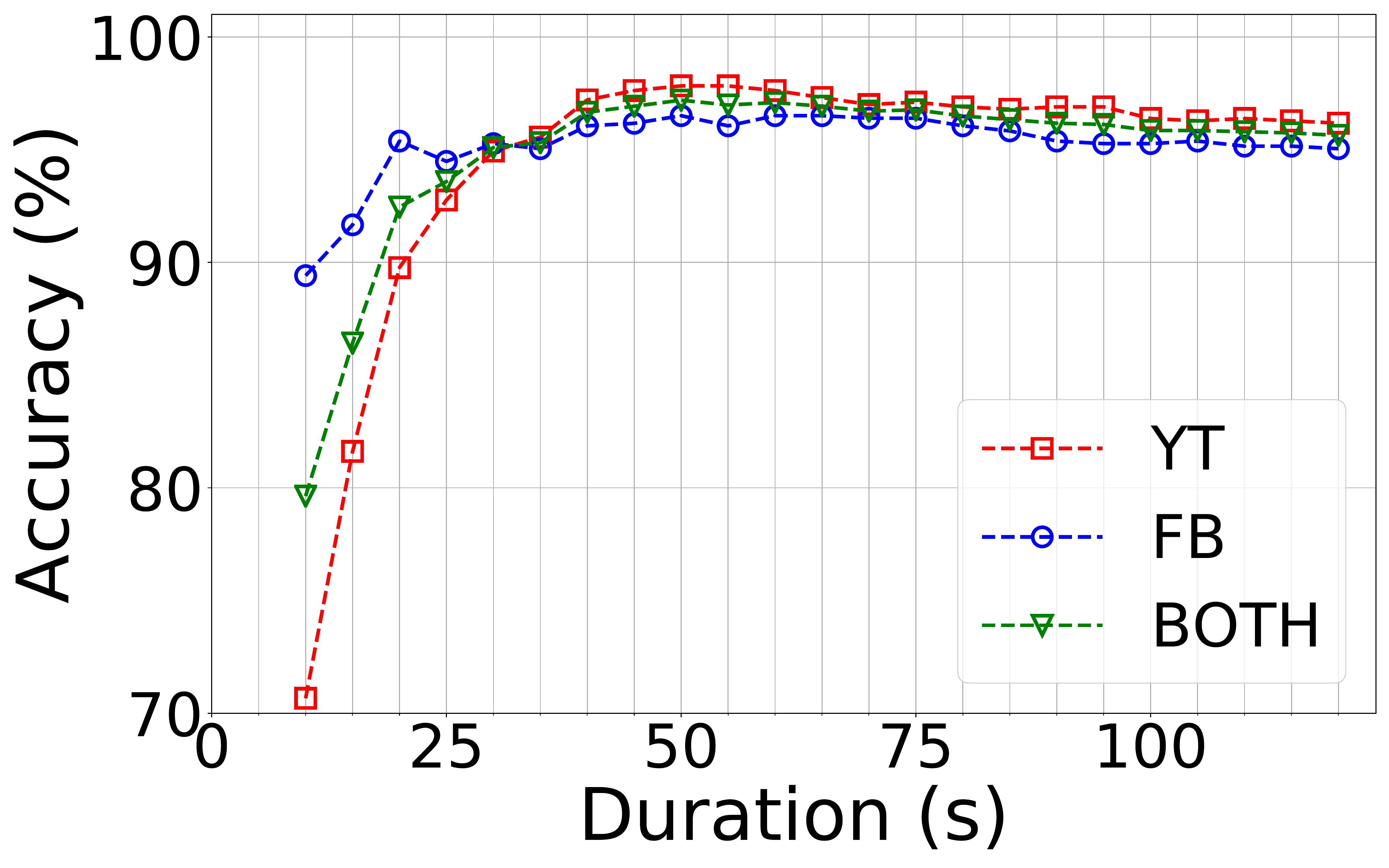}\vspace{-2mm}
    \caption{Accuracy by the video duration for \textsf{DS-pkt} data}
    \label{figure:near realtime performance}
  \end{subfigure}
  \hfill
  \begin{subfigure}{.49\columnwidth}
    \centering
    \includegraphics[width=\linewidth]{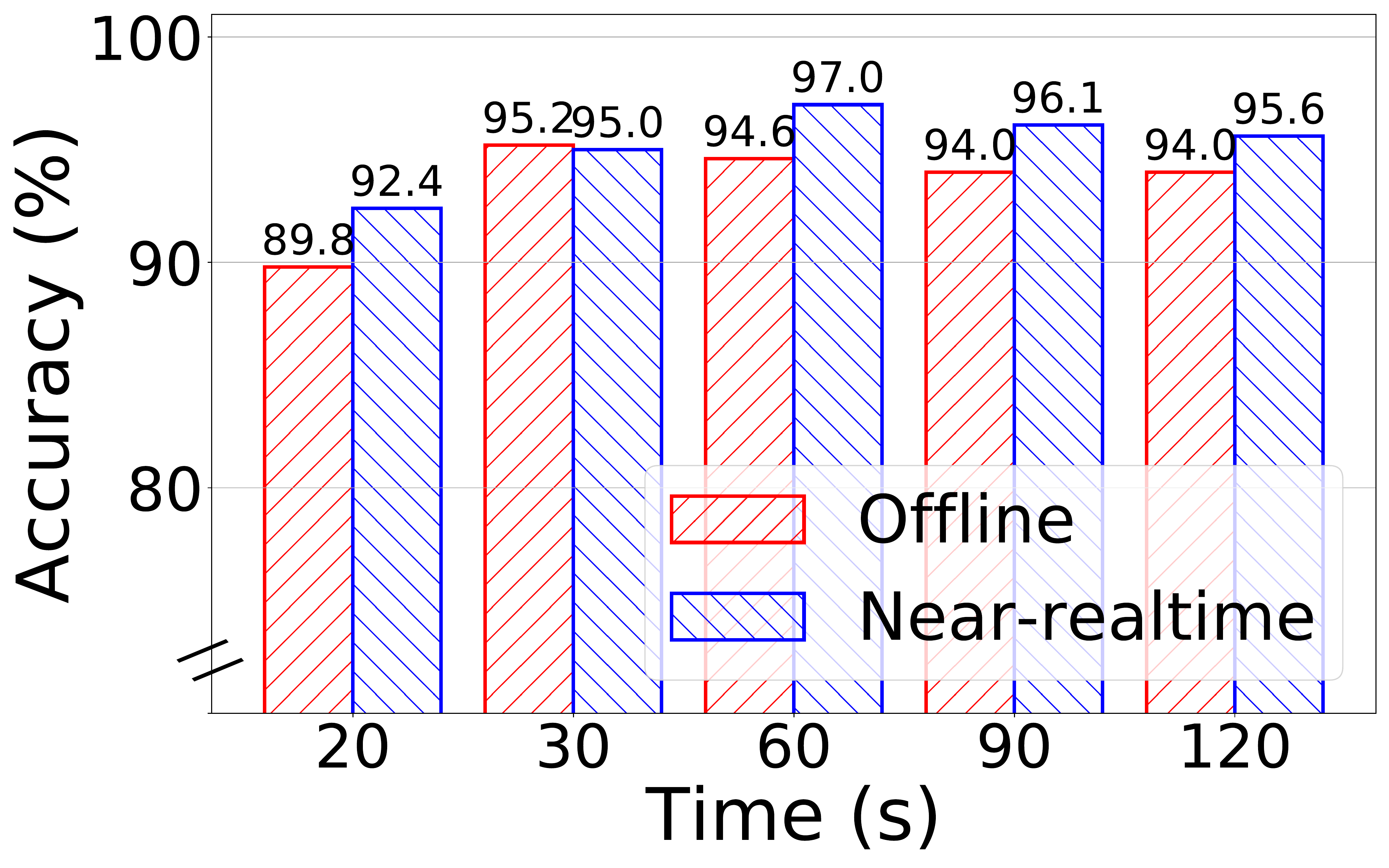}\vspace{-2mm}
    \caption{Comparison with offline model, selected duration: BOTH}
    \label{subfig:realtimevsoffline}
  \end{subfigure}\vspace{-4mm}
\caption{Near real-time model performance}
\vspace{-3mm}
\label{figure: packet level realtime performance}
\end{figure}

Figure~\ref{figure:near realtime performance} shows performance of our near real-time classification model. The model performance gradually increases until 30s, achieving (>95\%) accuracy and stabilizes after 50s. Note that, we observe a similar variation in F1 score for all three traffic types. Comparing with the results in Table \ref{table: performance non realtime packet level} at 20s, real-time model give 4.4\% (8.7\%), 1.7\% (2.8\%) and 2.9\% (5.6\%)  improvement in accuracy (F1 score) for YT, FB and BOTH respectively. Figure~\ref{subfig:realtimevsoffline} depicts that real-time model achieves better accuracy compared to offline. 

\emph{Overall, both offline and near real-time classification models are able to identify \ang{360} video flows with accuracy above 95\% with just 30s -- 50s of data packet stream, which enables ISPs and MNOs to deploy traffic shaping and policing solutions in near real-time. Flow level classification is more applicable for offline tasks such as network capacity planning and demand prediction, as the flow features can only be calculated after streaming the entire video.}

\vspace{-3mm}
\begin{figure}[h]
  \centering
  \captionsetup{justification=centering}
  \begin{subfigure}{.49\columnwidth}
    \centering
    \includegraphics[width=\linewidth]{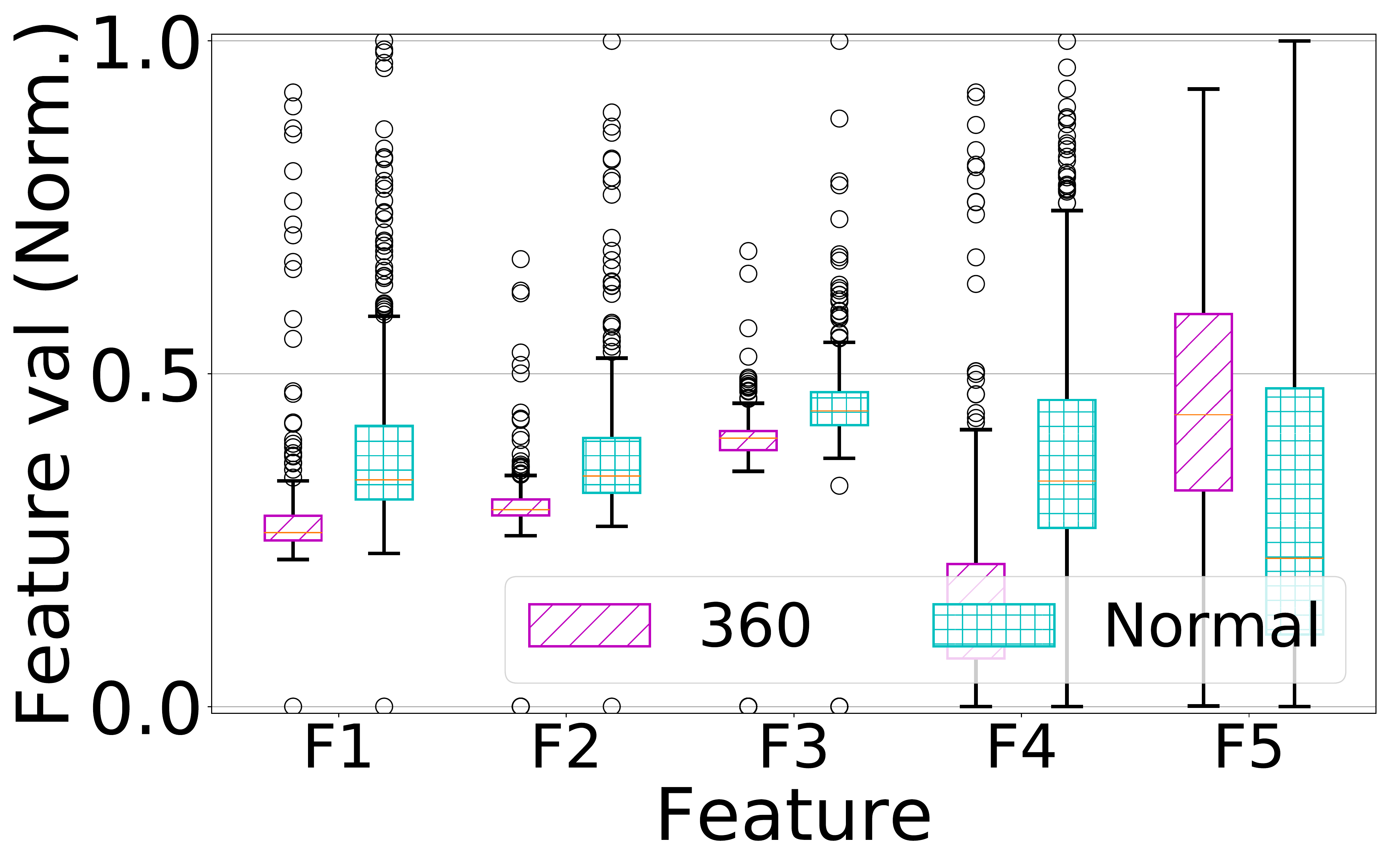}\vspace{-2mm}
    \caption{\textsf{DS-pkt: first 5 features}}
    \label{figure: important_feat_dist_DS_pkt}
  \end{subfigure}
  \hfill
  \begin{subfigure}{.49\columnwidth}
    \centering
    \includegraphics[width=\linewidth]{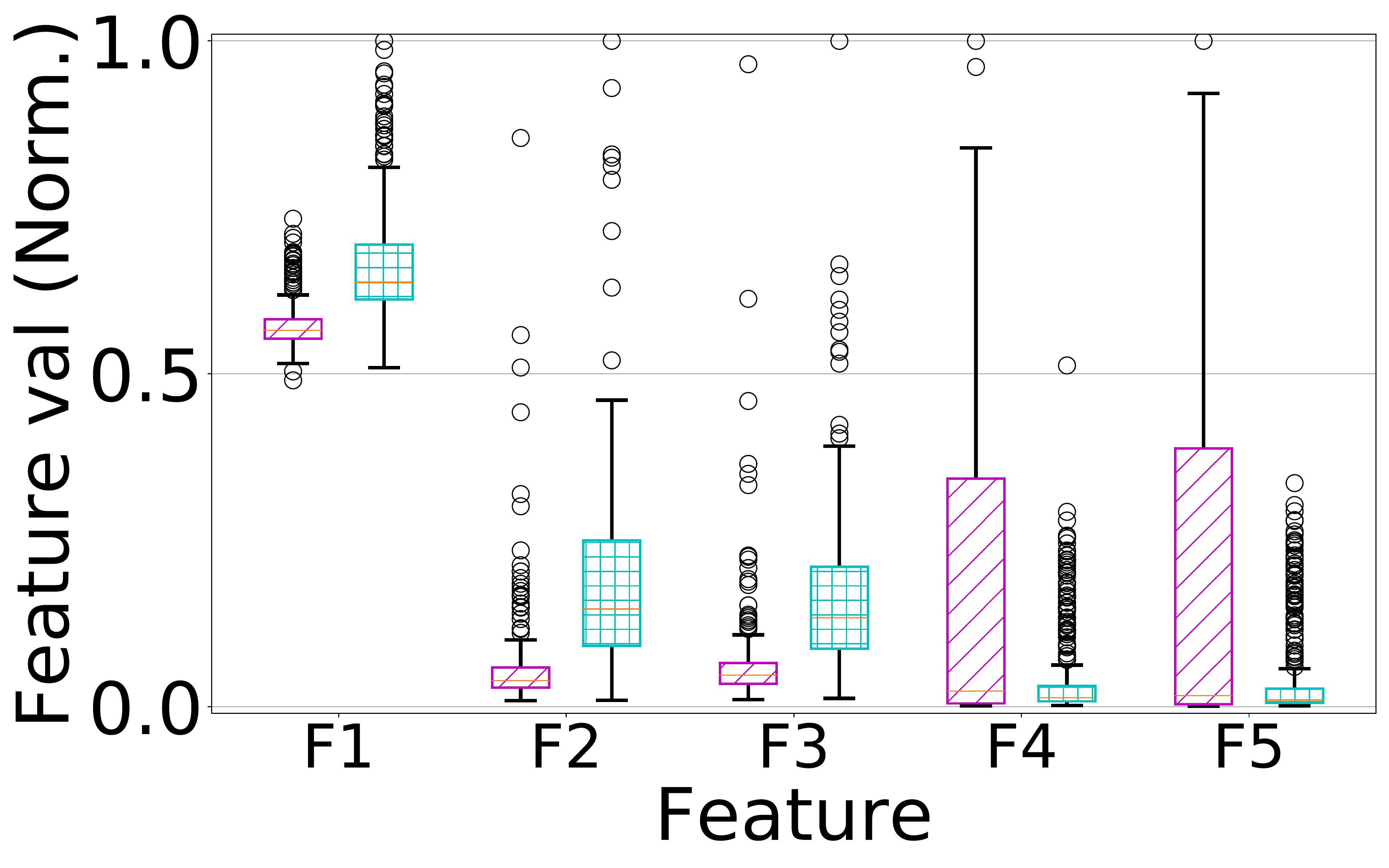}\vspace{-2mm}
    \caption{\textsf{DS-flw: first 5 features}}
    \label{figure: important_feat_dist_DS_flw}
  \end{subfigure}
  
  \begin{subfigure}{.49\columnwidth}
    \centering
    \includegraphics[width=\linewidth]{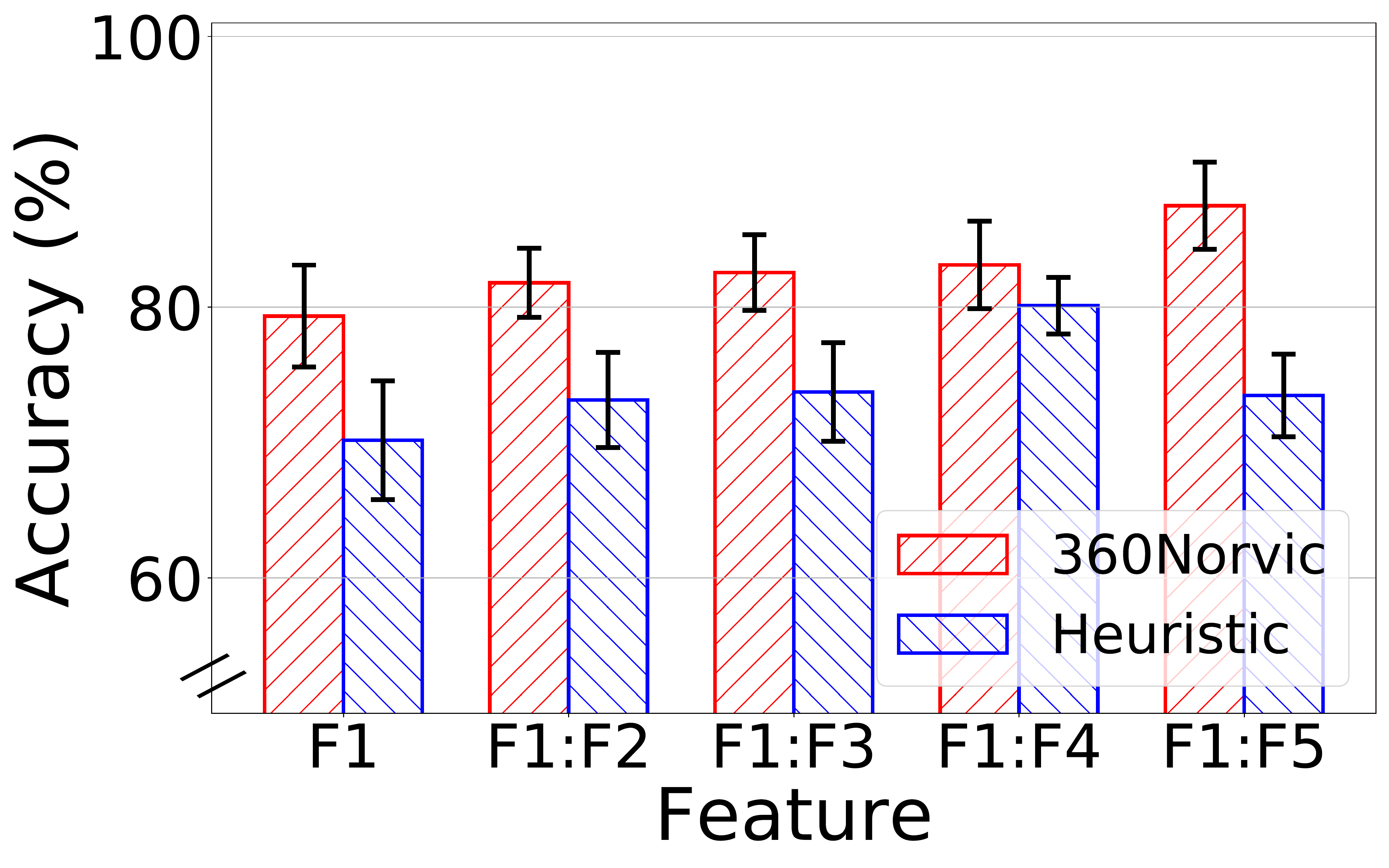}\vspace{-2mm}
    \caption{\textsf{DS-pkt: adding first 5 features cumulatively}}
    \label{figure:heuristic_ds_pkt}
  \end{subfigure}
  \hfill
  \begin{subfigure}{.49\columnwidth}
    \centering
    \includegraphics[width=\linewidth]{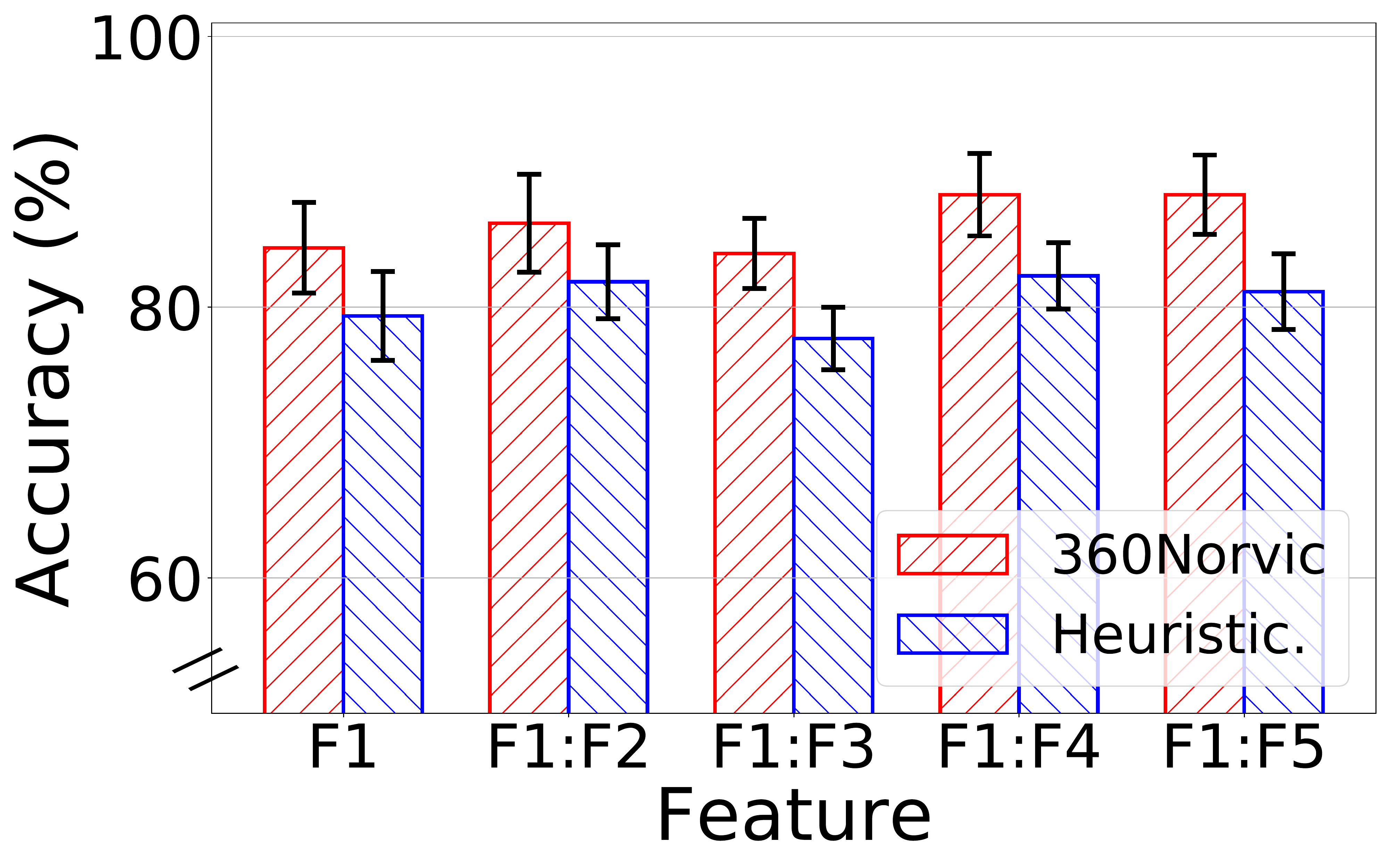}\vspace{-2mm}
    \caption{\textsf{DS-flw: adding first 5  features cumulatively}}
    \label{figure:heuristic_ds_flow}
  \end{subfigure}
  \vspace{-4mm}
\caption{Most important feature distribution
(\ref{figure: important_feat_dist_DS_pkt},\ref{figure: important_feat_dist_DS_flw}), Heuristic comparison
(\ref{figure:heuristic_ds_pkt},\ref{figure:heuristic_ds_flow}) for "BOTH" traffic type }
\label{figure: critical feature analysis}
\end{figure}

\vspace{-4mm}
\section{Analysis}
\label{sec: robustness of models}

\subsection{Critical features}
\label{feature_impact}
We observe that features related to the packet sizes are the most critical ones for \textsf{DS-pkt} scenario, while features related to burstiness and frame size become prominent in  \textsf{DS-flw} scenario.
Figure \ref{figure: important_feat_dist_DS_pkt} and \ref{figure: important_feat_dist_DS_flw}, shows the most important 5 feature value distributions  for 'BOTH' traffic type. The values between quartile~1 and quartile~3 always has a clear separation between \ang{360} and regular video. However, still there is a significant overlap comparing the entire range. Moreover, compared to FB, YT feature values are more identical to video type further confirming the higher classification accuracy for YT in Table \ref{table: performance non realtime packet level} (cf. Figure~\ref{figure:data dl_ul yt}~and~\ref{figure:data dl_ul fb}). 
We see that features such as throughput dl are less significant, but fall with in first 10 features, because they are highly  susceptible to deviated traffic behaviour due to the video content and network availability. Therefore, content independent uplink features contribute more towards classification.

\vspace{-2mm}
\subsection{Comparison with heuristic approach}
We compare \model{} to a heuristic approach considering the five most important features\footnote{F1~:~F$N$ denotes combined first $N$ number of features} in Figure~\ref{figure:heuristic_ds_pkt}~and~\ref{figure:heuristic_ds_flow}.
The heuristic uses a threshold per feature, and it distinguishes between \ang{360} and normal format using a majority rule on the per-feature classification. 
\model{} accuracy is always higher than the one of the heuristic approach, i.e.,  
the average accuracy difference is of 8.7($\pm$3.5)\% and 5.7($\pm$1.0)\% for \textsf{DS-pkt} and \textsf{DS-flw} respectively. We also observe additional features provides noticeable benefits to \model{} while the performance of the heuristic are not clearly improved.

\vspace{-2mm}
\subsection{Effect of video content \& its popularity}
\begin{figure}[h]
    \centering
    \includegraphics[width=0.9\linewidth]{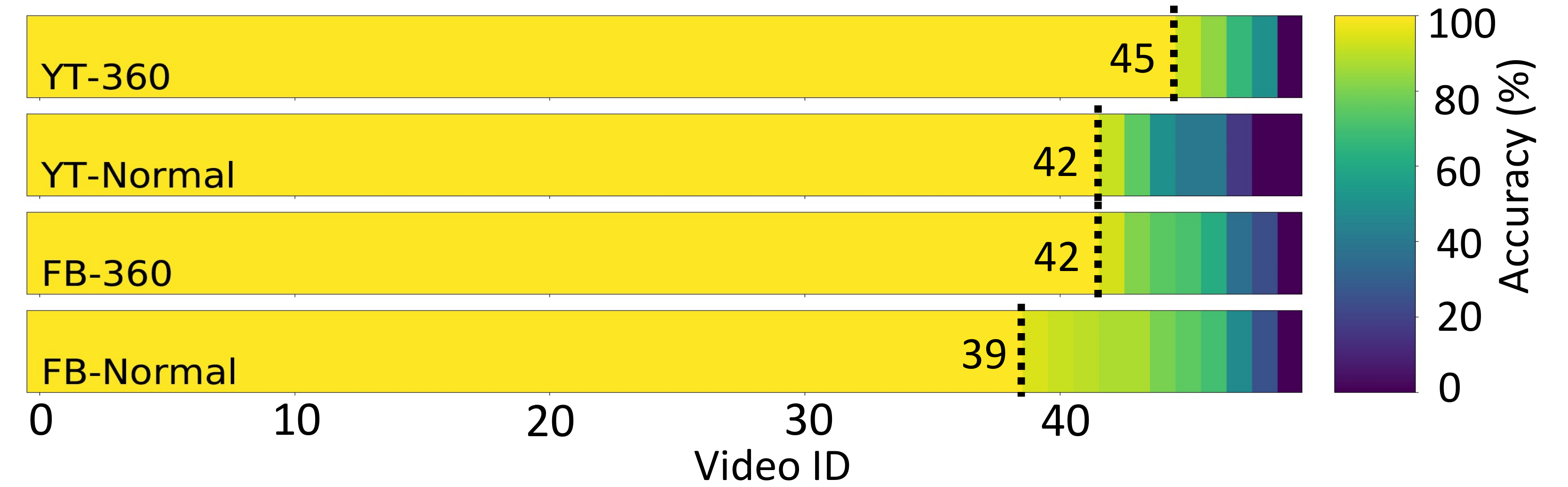}\vspace{-4mm}
  \caption{Avg. classification acc.
for each video from \textsf{DS-pkt}
  }
  \label{figure:effect of content}
\end{figure}

We collect individual classification accuracy for each video, streamed at the same conditions, from \textsf{DS-pkt}. Figure~\ref{figure:effect of content} shows the average accuracy for a video, as the same video is tested several times within 20 different train/test splits. In category, majority is classified with 100\% accuracy
(dotted line for 100\% accuracy margin).
Overall, out of 50 videos in each category, only 3 YT-\ang{360}, 5 YT-Normal, 4 FB-\ang{360} and 6 FB-Normal videos show less than 80\% accuracy. This shows that \model{} classification performance is content independent, supporting the generalization of our solution. 

In ML classifiers, we avoided traces of the same video appearing in train and test datasets (Section \ref{sec:Results and analysis}). However,
in general,
it can be assumed that most popular videos have already been seen by the ISP or MNO~\cite{cls,islam2013revisiting}.
We take 70\% of traces of each video as training data and remaining for testing. 
From the experimental results with \textsf{DS-pkt}, we could observe that accuracy of each platform was increased by 0.9-1.6\%, which can give huge impact in large scale networks. 

\vspace{-2mm}

\vspace{-6mm}
\begin{figure}[h]
  \centering
  \captionsetup{justification=centering}
   \begin{subfigure}{\columnwidth}
    \centering
    \includegraphics[width=\linewidth]{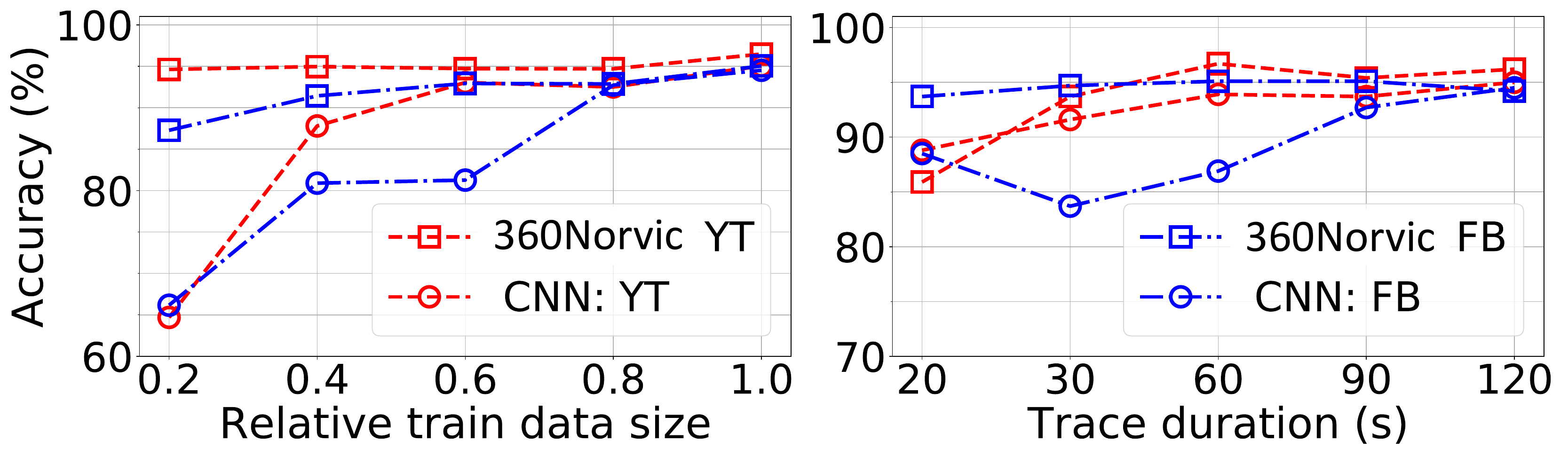}\vspace{-2mm}
  \caption{Comparison with CNN:(left)-limited training data (duration of the trace is 120s) \& (right)-trace duration(\# of traces are same)}
  \label{figure:cnn_vs_xgboost}
  \end{subfigure}
  
  \begin{subfigure}{\columnwidth}
    \centering
    \includegraphics[width=\linewidth]{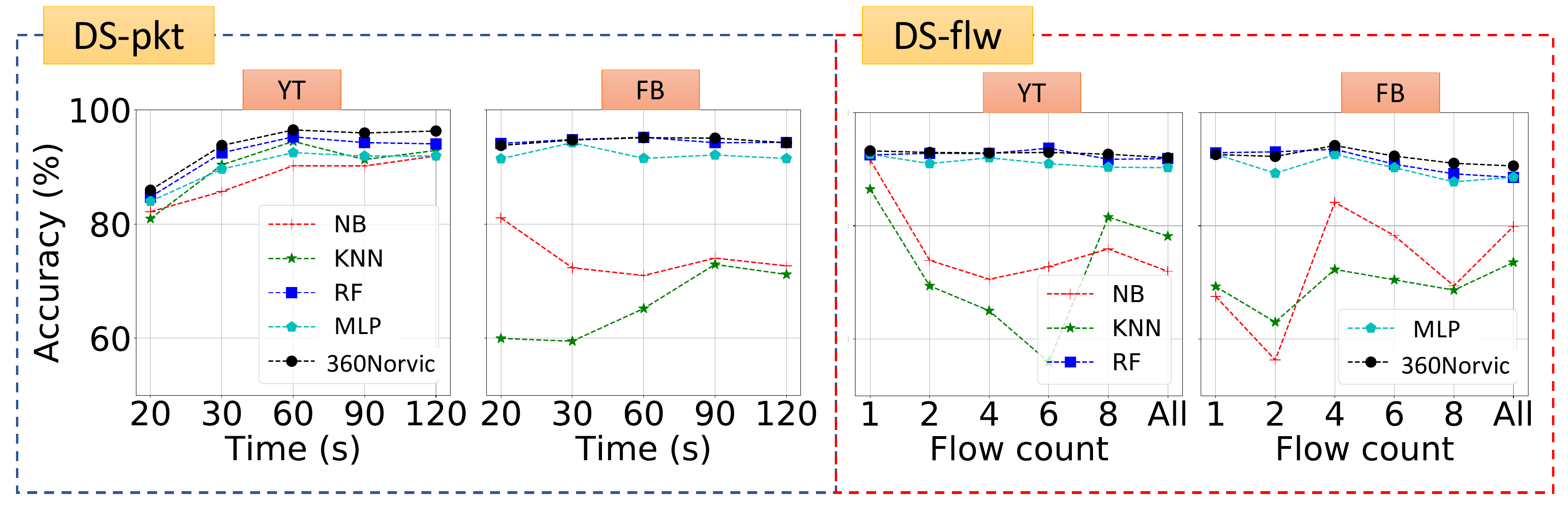}\vspace{-2mm}
  \caption{Comprison with other ML models for the configurations in Table~\ref{table: performance non realtime packet level}: \textsf{DS-pkt} (left) \& Table~\ref{table: performance offline flow level}: \textsf{DS-flw} (right)}
  \label{figure:traditional clf}
  \end{subfigure}
  \vspace{-4mm}
  \caption{Comparison with: CNN  \& Traditional clf}
  \label{figure:cnn_ml_vs_xgboost}
\end{figure}

\vspace{-5mm}
\subsection{Comparison with other ML models}
\label{subsec:models}

We compare \model{} with a modified Convectional Neural Network (CNN) architecture from~\cite{schuster2017beauty}, Multi-layer Perceptron (MLP)~\cite{deepcontent} and a set of selected ML classifiers used in literature\cite{velan2015survey,inferring_streaming_video_quality}. For the CNN, each video trace is sliced in bins of 0.25s; this  creates a feature vector with 480 samples for each trace of 120s long, as the input for the model. Figure~\ref{figure:cnn_vs_xgboost}~(left) reports the accuracy of \model{} (\textsf{DS-pkt}) and the CNN classifier for YT and FB as a function of the train set size and trace duration. When the entire train set is used (70\% of all traces),
value 1 in the left figure, \model{} performs as the CNN model. 
When reducing relative train data size, CNN accuracy drops by 30\% at most for both platforms, whereas the decrease for \model{} is only 1 (7)\% for YT (FB). Figure~\ref{figure:cnn_vs_xgboost}~(right) highlights that reducing the duration of the traces\footnote{When reducing duration the size of the input to the CNN model is also reduced proportionally.} the CNN accuracy for FB reduces to a minimum of 84\% at 30s. For YT, for both CNN and \model{} accuracy decrease steadily. 

Figure \ref{figure:traditional clf} reports the accuracy of \model{}, Naive Bayes (NB), k-nearest neighbors (KNN), Multi-layer Perceptron (MLP)~\cite{deepcontent} and Random Forest (RF) algorithms. Overall, both NB and KNN poorly performs in all cases, except for YT in packet level (\textsf{DS-pkt}). RF provides only slightly lower performance than \model{} while the MLP does not improve performance when compared to \model{}, while being more complex.
\emph{These results highlight that \model{} provides the best performance when compared to traditional ML algorithms, while using complex models (i.e., CNN, MLP) would not provide any benefits but rather hurt classification performance when there is not enough data available for training.}

\vspace{-1mm}
\section{Mobile Network Operator Pilot}
\label{sec:pilot}

We run a pilot in the commercial network of a large MNO to show the feasibility and effectiveness of \model~ in production settings. We first describe how \model~builds on the measurement infrastructure of the MNO to perform 360 video classification. Then, we present the methodology and results of the pilot, and discuss deployment considerations for large scale \model~usage by the MNO.
Note that for the pilot we only have access to the logs generated by a phone we controlled and no access to any information related to other customers of the MNO.

\noindent
\textbf{\model~in the MNO infrastructure.}
As depicted in Figure~\ref{figure:360_norvic_overview}~\textcircled{\footnotesize3}{},
\model~leverages flow level logs collected at the transparent proxy the MNO deploys. The proxy aims to optimize the entire mobile traffic, and to log performance metrics about users' transactions. A  transaction is an entry in the monitoring logs generated by the middlebox, and corresponds to an individual flow handling encrypted or non-encrypted traffic generated by an app of a mobile phone connected to the MNO. 
Note that flow-level statistics are computed on the entire flow duration. From all flows, \model~identifies the ones related to video sessions by using some of the information reported by the middlebox, (e.g., video flow tag indicating whether the flow carries video traffic, amount of bytes transferred within the flow), and computes the required feature vectors. As statistics over flows are already computed by the middlebox, the flow selection process does not generate significant overhead, and can be performed in near real time. Note that these flow data can be slightly different from \textsf{DS-flw}, due to the differences in processing mechanisms used by the middle box and Wireshark tool used to generate \textsf{DS-flw}.
Finally, \model~classification step is divided in two phases, namely \emph{Training} and \emph{Inference}. In the former, we train and validate the model on a set of flows generated via controlled experiments, while in the latter we apply the trained model (i.e., model inference) to all flows of the MNO customers in the wild. Note that only the inference step has to be performed near real time, and this operation is indeed performed in the order of few hundreds of milliseconds.   

\begin{figure}[h]
    \centering
    \includegraphics[width=\linewidth]{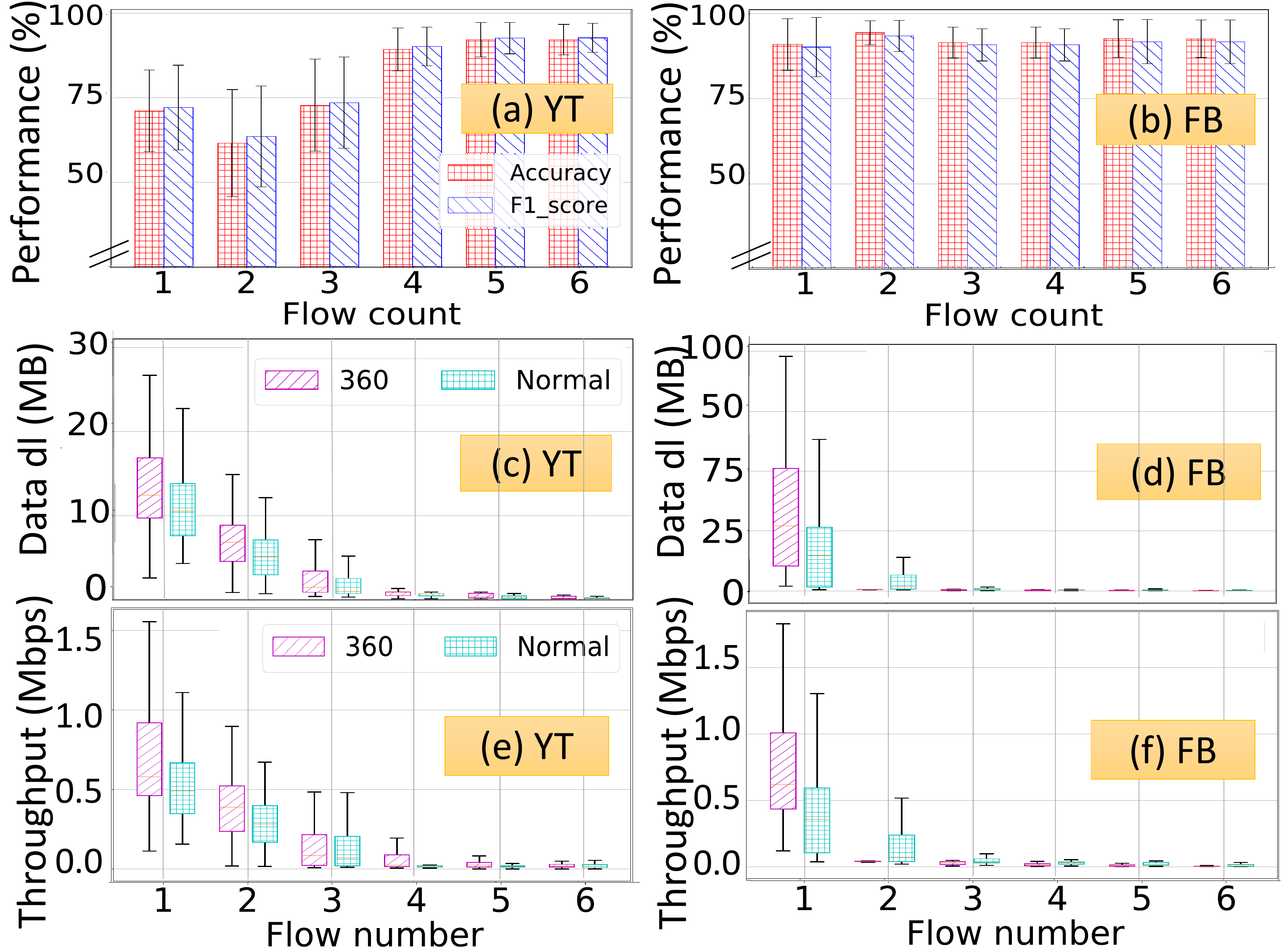}\vspace{-4mm}
    \caption{DS-mno analysis for FB and YT: (a,b)--Acc. \& F1-Score by \# of flows, (c,d)--Data dl  and (e,f)--Throughput DL distribution by the flow sorted according to bytes dl value.}
    \label{figure:combined_analysis_mno}
    \vspace{-4mm}
\end{figure}

\noindent
\textbf{MNO-pilot overview and results}
The pilot consists in one instance of the \emph{Training} phase.
Following the methodology in Sec.~\ref{sec:methodology}, we play 100 normal video traces and 100 \ang{360} video traces from each platform, YT \& FB, with a Xiaomi Redmi 4A smartphone connected to the 4G network of the MNO. 
We randomly select videos from ~\cite{ytVR,ytTrend,Mostpopular360vidFB,360videosFB,vtrnd,unilad}.
We name this controlled dataset as \textsf{DS-mno} and we analyze it offline.
\model~ achieves an average accuracy of 92.0\% ($\pm 5.1\%$) and average F1 score of 92.6\% ($\pm 4.6\%$) for YT. Similarly, for FB, the accuracy is 94.4\% ($\pm 3.5\%$) and F1 score is 93.4\% ($\pm 4.5\%$). 

Figure~\ref{figure:combined_analysis_mno}a~and~~\ref{figure:combined_analysis_mno}b reports \model~accuracy as a function of the number of flows used for classification. Note that flows are sorted in decreasing order of bytes downloaded as detailed in 
Figure~\ref{figure:combined_analysis_mno}c~and~~\ref{figure:combined_analysis_mno}d.
For FB, we observe that the usage of even a single flow for classification results in high accuracy. Indeed, a single flow carries the large majority of the video content 
(cf. Figure~\ref{figure:combined_analysis_mno}d.) 
and it is sufficient to discriminate between 360 and regular videos.  
Conversely, for YT we need at minimum four flows to achieve high accuracy. Indeed, video content is spread across multiple flows, as shown in Figure~\ref{figure:combined_analysis_mno}c, where the first four flows above 1 MB carry the majority of the content. Finally, the most important features used by the model for classification in the pilot are similar to the ones highlighted in Section~\ref{feature_impact}, especially the downlink
throughput, reported for YT and FB in Figure~\ref{figure:combined_analysis_mno}e~and~\ref{figure:combined_analysis_mno}f. Similar to the previous analysis, first six and three flows present distinguishable values of the considered feature for YT and FB respectively.

\emph{On the one hand, these results show the effectiveness and feasibility of \model~using data collected in a commercial network. On the other hand, they are in line with what is observed in the rest of the paper in the experimental testbed.}

\vspace{-1mm}
\section{Conclusion and future work}

This paper presents \model, a ML-based YT \& FB \ang{360} video traffic classification engine to identify \ang{360} video traffic, taking both packet and flow level data as the input to provide near-realtime and offline predictions. For the packet level data, we achieve (>95\%) accuracy for both near-realtime and offline classification, whereas for aggregated flows, \model~ overall provides (>92\%) accuracy. 
Further, we analyse effect of video content and its popularity on classification and compare \model{} with other ML and DNN models. Finally, we demonstrate the practical feasibility of \model~by conducting a pilot in a MNO for Facebook and YouTube \ang{360} video classification, which exceeds 92\% accuracy. 
In future work, we aim to conduct classification on viewport-aware streaming to see its impact on \model{} performance. Also, we plan to extend our pilot study to a large-scale data gathering in the wild to demonstrate benefits to MNOs such as to explore usage of \ang{360} videos.

\bibliographystyle{ACM-Reference-Format}
\balance
\newpage
\bibliography{main}

\end{document}